\documentclass[iop]{emulateapj}

\usepackage{natbib}
\usepackage{amsmath, amssymb}
\usepackage{times}
\usepackage{apjfonts, graphicx, graphics}
\usepackage[breaklinks,colorlinks,urlcolor=blue,citecolor=blue]{hyperref}
\usepackage[all]{hypcap}
\usepackage{aas_macros}

\newcommand{\Mpc}{\rm\; Mpc}
\newcommand{\kpc}{\rm\; kpc}
\newcommand{\pc}{\rm\; pc}
\newcommand{\km}{\rm\; km}
\newcommand{\m}{\rm\; m}
\newcommand{\cm}{\rm\; cm}

\newcommand{\cmsq}{\hbox{$\cm^2\,$}}

\newcommand{\yr}{\rm\; yr}
\newcommand{\Gyr}{\rm\; Gyr}

\newcommand{\s}{\rm\; s}

\newcommand{\GHz}{\rm\; GHz}
\newcommand{\MHz}{\rm\; MHz}

\newcommand{\K}{\rm\; K}

\newcommand{\Msun}{\hbox{$\rm\thinspace M_{\odot}$}}

\newcommand{\Msunpyr}{\hbox{$\Msun\yr^{-1}\,$}}

\newcommand{\keV}{\rm\; keV}

\newcommand{\erg}{\rm\; erg}
\newcommand{\Jy}{\rm\; Jy}
\newcommand{\mJy}{\rm\; mJy}

\newcommand{\ergpcmcu}{\hbox{$\erg\cm^{-3}\,$}}

\newcommand{\ergps}{\hbox{$\erg\s^{-1}\,$}}

\newcommand{\kmps}{\hbox{$\km\s^{-1}\,$}}

\newcommand{\kmpspMpc}{\hbox{$\kmps\Mpc^{-1}\,$}}

\newcommand{\Zsun}{\hbox{$\thinspace \mathrm{Z}_{\odot}$}}

\newcommand{\pcmsq}{\hbox{$\cm^{-2}\,$}}
\newcommand{\pcmcu}{\hbox{$\cm^{-3}\,$}}

\providecommand{\e}[1]{\ensuremath{\times 10^{#1}}}

\newcommand{\Jykmps}{\hbox{$\Jy\km\s^{-1}\,$}}
\newcommand{\Kkmps}{\hbox{$\K\km\s^{-1}\,$}}
\newcommand{\Kkmpspcsq}{\hbox{$\K\km\s^{-1}\pc^{2}\,$}}

\newcommand{\acounits}{\hbox{$\Msun(\Kkmpspcsq)^{-1}$}}

\shorttitle{CO-to-H$_2$ Conversion Factor in BCGs}
\shortauthors{Vantyghem et al.}

\begin{document}

\title{A $^{13}$CO Detection in a Brightest Cluster Galaxy}
\author{A.~N. Vantyghem$^{1}$}
\author{B.~R. McNamara$^{1, 2}$}
\author{A.~C. Edge$^3$}
\author{F. Combes$^{4, 5}$}
\author{H.~R. Russell$^{6}$}
\author{A.~C. Fabian$^{6}$}
\author{M.~T. Hogan$^{1, 2}$}
\author{M. McDonald$^{7}$}
\author{P.~E.~J. Nulsen$^{8, 9}$}
\author{P. Salom{\'e}$^{4}$}

\affil{
    $^1$ Department of Physics and Astronomy, University of Waterloo, Waterloo, ON N2L 3G1, Canada; \href{mailto:a2vantyg@uwaterloo.ca}{a2vantyg@uwaterloo.ca} \\ 
    $^2$ Perimeter Institute for Theoretical Physics, Waterloo, Canada \\
    $^3$ Department of Physics, Durham University, Durham DH1 3LE \\
    $^4$ LERMA, Observatoire de Paris, CNRS, UPMC, PSL Univ., 61 avenue de l'Observatoire, 75014 Paris, France \\
    $^5$ Coll{\`e}ge de France, 11 place Marcelin Berthelot, 75005 Paris \\
    $^6$ Institute of Astronomy, Madingley Road, Cambridge CB3 0HA \\
    $^7$ Kavli Institute for Astrophysics and Space Research, Massachusetts Institute of Technology, 77 Massachusetts Avenue, Cambridge, MA 02139, USA \\
	$^8$ Harvard-Smithsonian Center for Astrophysics, 60 Garden Street, Cambridge, MA 02138, USA \\
    $^9$ ICRAR, University of Western Australia, 35 Stirling Hwy, Crawley, WA 6009, Australia
}

\begin{abstract}

We present ALMA Cycle 4 observations of CO(1-0), CO(3-2), and $^{13}$CO(3-2) line emission in the brightest cluster galaxy of RXJ0821+0752. This is one of the first detections of $^{13}$CO line emission in a galaxy  cluster. Half of the CO(3-2) line emission originates from two clumps of  molecular gas that are spatially offset from the galactic center. These  clumps are surrounded by diffuse emission that extends $8~{\rm kpc}$ in length. The detected $^{13}$CO emission is confined entirely to the two bright  clumps, with any emission outside of this region lying below our detection  threshold. Two distinct velocity components with similar integrated fluxes are  detected in the $^{12}$CO spectra. The narrower component ($60~{\rm km}~{\rm s}^{-1}$ FWHM) is  consistent in both velocity centroid and linewidth with $^{13}$CO(3-2) emission,  while the broader ($130-160~{\rm km}~{\rm s}^{-1}$), slightly blueshifted wing has no associated $^{13}$CO(3-2) emission. A simple local thermodynamic model indicates that the $^{13}$CO  emission traces $2.1\times 10^{9}~{\rm M}_\odot$ of molecular gas. Isolating the $^{12}$CO velocity component that accompanies the $^{13}$CO emission yields a CO-to-H$_2$  conversion factor of $\alpha_{\rm CO}=2.3~{\rm M}_{\odot}~({\rm K~km~s^{-1}})^{-1}$, which is a factor of two  lower than the Galactic value.  Adopting the Galactic CO-to-H$_2$ conversion factor in brightest cluster  galaxies may therefore overestimate their molecular gas masses by a factor of two. This is within the object-to-object scatter from extragalactic sources, so calibrations in a larger sample of clusters are necessary in order to confirm  a sub-Galactic conversion factor.

\end{abstract}


\section{Introduction}

The brightest cluster galaxies (BCGs) at the centers of some galaxy clusters 
are rich in molecular gas \citep{Edge01, Salome03}. Gas-rich BCGs also host 
luminous emission-line nebulae \citep{Heckman89, Crawford99} and star formation 
rates rivaling those of starburst galaxies \citep{mcn94, ODea08, Tremblay15}. 
These cooling signatures are observed preferentially when the central cooling 
time of the hot intracluster gas falls below $1\Gyr$ \citep{Rafferty08}, or 
when the central entropy lies below $30\keV\cmsq$ \citep{Cavagnolo08}. The 
implication is that the molecular gas in BCGs is formed from the cooling of 
the hot atmosphere.

Uninhibited cooling of the hot atmosphere would produce far more cold gas 
and star formation than is observed. Instead, the rate of cooling is regulated 
by active galactic nucleus (AGN) feedback (see \citealt{mcn07, mcn12, Fabian12} 
for reviews). Radio jets launched by the AGN inflate buoyant radio bubbles  
(X-ray cavities), drive shocks, and generate sound waves, heating the gas 
throughout the cluster core \citep[e.g.][]{mcn00, Blanton01, Fabian06}.
The power output by the AGN is correlated with the cooling rate of the cluster 
gas in a large sample of groups and clusters \citep{Birzan04, Dunn06, Rafferty06}, 
indicating that AGN are capable of preventing the bulk of the hot gas from cooling. 
Residual cooling from the hot atmosphere can then form the observed molecular 
gas reservoirs. Accretion of this gas onto the AGN likely fuels AGN feedback, 
establishing a feedback loop by connecting gas cooling to heating \citep{Pizzolato05, 
Gaspari13, Li14a}.

A growing body of evidence indicates that AGN feedback plays a direct role 
in shaping the distribution of molecular gas. In the Perseus cluster, 
filamentary H$\alpha$ emission extends radially from the BCG \citep{Conselice01, 
Hatch06}, with two prominent filaments oriented towards an X-ray cavity 
\citep{Fabian03}. The nebular emission is coincident with soft X-rays, 
molecular hydrogen \citep{Lim12}, and CO emission \citep{Salome06, Salome11}. 
Recent ALMA observations of BCGs have also revealed molecular filaments that 
trail X-ray cavities \citep{mcn14, Russell16, Russell17, Vantyghem16}. 
This gas has either been lifted directly from the cluster core by buoyantly 
rising radio bubbles, or it has cooled in situ from hot gas that has been 
uplifted to an altitude where it becomes thermally unstable \citep{Revaz08, mcn16}. 
These gas flows comprise a significant fraction of the total molecular 
gas mass.
Redshifted absorption line measurements in other systems imply that molecular 
clouds are falling toward the central black hole, perhaps indicating that the 
clouds return to the central galaxy in a circulation flow \citep{David14, 
Tremblay16}.

The molecular gas in BCGs is observed predominantly using CO\footnote{When 
referring to individual transitions, we refer to the most common $^{12}$C$^{16}$O 
isotopologue as simply CO.} 
as a tracer molecule. Converting the measured intensity of the CO rotational line 
into a molecular gas mass requires the assumption of a CO-to-H$_2$ 
conversion factor. This conversion factor, also referred to as the X-factor, 
is defined as the ratio between H$_2$ column density, $N({\rm H}_2)$ in $\pcmsq$, 
and integrated intensity of the CO(1-0) line, $W({\rm CO})$ in $\Kkmps$:
\begin{equation}
  N({\rm H}_2) = X_{\rm CO} W({\rm CO}).
  \label{eqn:Xco}
\end{equation}
In the Milky Way and nearby spiral galaxies $X_{\rm CO}$ is calibrated to 
be $2\e{20}\pcmsq(\K\kmps)^{-1}$ \citep[for a review, see][]{Bolatto13}. 
However, the Galactic $X_{\rm CO}$ is not universal, and
independent calibrations within BCGs are not available. The standard 
practice has been to adopt the Galactic value with a factor of two 
uncertainty.
This approach is generally justified by the near-solar metallicities 
at the centers of galaxy clusters. The molecular clouds are also expected 
to be cold \citep[$\ll100\K$;][]{Ferland94}, resembling those in the Galaxy. The 
linewidths of the individual clouds seen in absorption ($\sim5\kmps$) 
are similar to giant molecular clouds, further indicating that the cold 
gas in BCGs resembles Galactic clouds.

Significant deviations from the Galactic $X_{\rm CO}$ are observed in 
ultra-luminous infrared galaxies (ULIRGs) and jet-driven outflows. 
The physical conditions of the molecular gas in ULIRGs, which exhibit 
extreme star formation, differ greatly from those in the disks 
of normal galaxies. The gas is located in an extended warm phase with 
volume and column densities that are much higher than in normal disks 
\citep[e.g.][]{Jackson95, Ward03}. This leads to overluminous CO 
emission, reducing the standard CO-to-H$_2$ conversion factor to 
$X_{\rm CO} = 0.4\e{20} \pcmsq(\K\kmps)^{-1}$ \citep{Downes98}.

Massive outflows of molecular gas can be driven by intense radiation 
or radio jets \citep{Morganti05, Nesvadba06, Feruglio10, Rupke11, 
Sturm11, Alatalo11, Dasyra11, Cicone14, Tadhunter14, Morganti15}. 
The high CO (4-3)/(2-1) ratio in IC~5063 implies that the molecular 
gas along the jet-driven outflow is optically thin \citep{Dasyra16}. 
As a result, the CO-to-H$_2$ conversion factor may be reduced in 
these systems by an order of magnitude. Simulations of molecules forming 
along fast outflows powered by quasar-driven winds indicate that the 
conversion factor should be 25 times lower than Galactic \citep{Richings17}.

In this work we present an ALMA cycle 4 observation of the CO(1-0), CO(3-2), 
and $^{13}$CO(3-2) emission lines in the BCG of the RXJ0821+0752 galaxy 
cluster. This represents one of the first detections of $^{13}$CO within 
a BCG. 
Previous observations of NGC~1275 in the Perseus cluster detected both 
the $^{13}$CO(2-1) and $^{13}$CO(3-2) lines \citep{Bridges98}, while 
observations of A1835 and A1068 provided only upper limits \citep{Edge01}.
Due to the lower abundance of $^{13}$CO relative to $^{12}$CO, the $^{13}$CO 
emission is generally optically thin. 
As a result, the $^{13}$CO emission traces the full volume of its emitting 
region, allowing a direct measure of its column density. We use this to 
estimate the total H$_2$ column density, and by extension the molecular 
gas mass. This provides an estimate of the CO-to-H$_2$ conversion factor 
that can be compared to the Galactic value in order to evaluate previous 
mass estimates of BCGs.

RXJ0821+0752 is a cool core galaxy cluster with a brightest cluster galaxy
that contains one of the largest cold gas reservoirs known \citep{Edge01}. 
Its luminous CO emission corresponds to a molecular gas mass of $10^{10}\Msun$, 
assuming the standard Galactic CO-to-H$_2$ conversion factor. Despite the 
large CO luminosity, no significant 1-0 S series of H$_2$ has been detected. 
Observations of the CO(1-0) and CO(2-1) lines showed that the molecular gas 
traces the H$\alpha$ emission, but is not centered on the galaxy \citep{Salome04}.
The IR luminosity of $8.47\e{44}\ergps$ corresponds to a star formation rate 
of $37\Msunpyr$ \citep{Quillen08, ODea08}.
The BCG also hosts significant quantities of dust, with a dust mass of 
$2.2\e{7}\Msun$ assuming a dust temperature of $40\K$ \citep{Edge01}.
Unlike other BCGs, the optical emission-line ratios resemble those of H{\sc ii} 
regions instead of AGN-dominated regions \citep{Crawford99}.

Throughout this paper we assume a standard $\Lambda$-CDM cosmology with 
$H_0=70\kmpspMpc$, $\Omega_{{\rm m}, 0}=0.3$, and $\Omega_{\Lambda, 0}=0.7$.
At the redshift of RXJ0821+0752 \citep[$z=0.109$;][]{Crawford95}, the angular 
scale is $1''=2.0\kpc$ and the luminosity distance is $510\Mpc$.

\section{Observations and Data Reduction}
\label{sec:obs}

The brightest cluster galaxy of the RXJ0821+0752 galaxy cluster (RA: 08:21:02.258, 
Dec: +07:51:47.28) was observed by ALMA Band 3 on 2016 Oct 30 and Nov 4 and Band 
7 on 2016 Oct 1 (Cycle 4, ID 2016.1.01269.S, PI McNamara). These observations covered 
the redshifted CO(1-0) and CO(3-2) lines at $103.848\GHz$ and $311.528\GHz$, 
respectively. An additional baseband in the Band 7 observation also covered the 
$^{13}$CO(3-2) line at $297.827\GHz$. The remaining three Band 3 basebands (91.857, 
93.732, and 105.733 GHz) and two Band 7 basebands (299.554 and 309.680 GHz) 
were used to measure the sub-mm continuum emission.
The observations used a single pointing centered on the BCG nucleus with a primary 
beam of 60 arcsec at CO(1-0) and 20 arcsec at CO(3-2).
The total on-source integration time was 86.7 minutes at CO(1-0) and 22.7 minutes 
at CO(3-2), each split into $\sim6$ minute long observations and interspersed 
with observations of the phase calibrator.
The observations used 40 antennas with baselines ranging from $18-1124\m$ for Band 3 
and $15-3247\m$ for Band 7.
The frequency division correlator mode was used for the CO(1-0) and CO(3-2) spectral 
line observations, providing a 1.875 GHz bandwidth with 488 kHz frequency resolution. 
This corresponds to a velocity resolution of $1.4\kmps$ at CO(1-0) and $0.47\kmps$ 
at CO(3-2), although the data were binned to a coarser velocity channels for subsequent 
analysis. The remaining basebands were configured with the time division correlator 
mode with a bandwidth of 2 GHz and frequency resolution of $15.625\MHz$. For the 
$^{13}$CO(3-2) line, this corresponds to a velocity resolution of $15.7\kmps$. 

The observations were calibrated in {\sc casa} version 4.7.0 \citep{casa} using the 
pipeline reduction scripts. Continuum-subtracted data cubes were created using 
{\sc uvcontsub} and {\sc clean}. Additional phase self-calibration could not be 
performed because of the very faint nuclear continuum source. The calibration of 
the water vapour radiometers failed for the CO(3-2) observation, so some streaks 
associated with the phase calibration remain in the image.
Images of the line emission were reconstructed using Briggs weighting with a robust 
parameter of 2. An additional $uv$ tapering was used to smooth the CO(3-2) and 
$^{13}$CO(3-2) images on scales below 0.1 arcsec. 
The final CO(1-0) data cube had a synthesized beam of 
$0.61''\times 0.59''$ (P.A. $-70.4^{\circ}$), and the CO(3-2) data cube was smoothed 
to match the $^{13}$CO(3-2) synthesized beam of $0.21''\times 0.165''$ (P.A. 
$37.2^{\circ}$). The CO(1-0), CO(3-2), and $^{13}$CO(3-2) images were binned to 
$3$, $5$, and $16\kmps$ velocity channels, respectively. All images were centered 
around the line emission, which corresponds to a redshift of $z=0.109$.
The RMS noise in the line-free channels were $0.5$, $1.1$, and 
$0.3\mJy~{\rm beam}^{-1}$, respectively.
Images of the continuum were created by combining line-free spectral channels from 
each baseband. No continuum emission was detected from the nucleus of the BCG. 
Instead, the continuum emission is coincident with the molecular gas, which is 
consistent with the structure at $1.4$ and $5\GHz$ \citep{BayerKim02}.

\section{Results}

\begin{figure}
  \centering
  \includegraphics[width=\columnwidth]{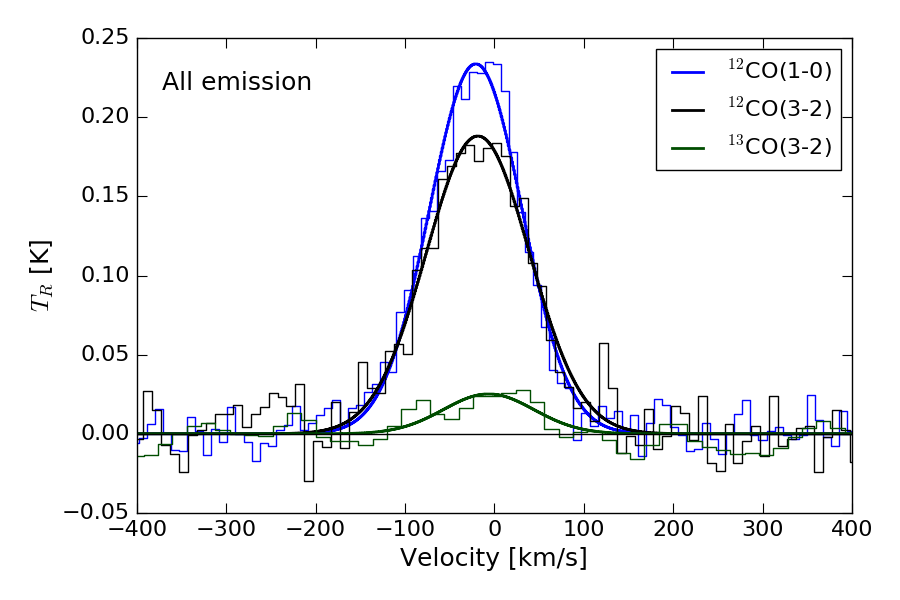}
  \caption{
    CO(1-0), CO(3-2), and $^{13}$CO(3-2) spectra extracted 
    from a $6.8''\times 5''$ ($13.6\times 10\kpc$) box encompassing all 
    of the line emission. The CO(1-0) and CO(3-2) spectra 
    have been smoothed to $9\kmps$ and $10\kmps$, respectively.
    The spatial resolution of the CO(3-2) and $^{13}$CO(3-2) images were 
    degraded to match the CO(1-0) resolution prior to the spectral extraction. 
  }
  \label{fig:isotopes}
\end{figure}

\subsection{Spectra}
\label{sec:spectra}

\begin{table*}
\begin{minipage}{\textwidth}
\caption{Spectral Fitting}
\begin{center}
\begin{tabular}{l c c c c c c}
\hline
\hline
Line   & Velocity Center & FWHM & $S_{\rm CO}\Delta v$ & $W({\rm CO})$ & $T_{\rm peak}$ & $L'_{\rm CO}/10^9$ \\
       & $\kmps$       & $\kmps$ & $\Jykmps$           & $\K\kmps$     & $\K$       & $\Kkmpspcsq$    \\
\hline
\multicolumn{7}{l}{{\bf All emission}} \\
CO(1-0) & $-20.7\pm0.6$ & $122.1\pm1.4$ & $8.06\pm0.08$ & $30.5\pm0.3$ & $0.235\pm0.004$ & $4.61\pm0.05$ \\
CO(3-2) & $-18.3\pm1.5$ & $136.1\pm3.6$ & $65.6\pm1.5$ & $27.3\pm0.6$ & $0.189\pm0.006$ & $4.17\pm0.10$ \\
$^{13}$CO(3-2) & $-6.4\pm8.2$ & $119\pm19$ & $7.0\pm1.0$ & $3.2\pm0.4$ & $0.025\pm0.005$ & $0.49\pm0.07$ \\
\multicolumn{7}{l}{{\bf Region tracing $^{13}$CO emission}} \\
CO(1-0) & $3.2\pm0.6$ & $61.7\pm2.0$ & $1.5\pm0.1$ & $68.5\pm5.4$ & $1.04\pm0.09$ & $0.86\pm0.06$ \\
               & $-19.2\pm1.3$ & $127.2\pm2.2$ & $3.0\pm0.1$ & $137.7\pm5.4$ & $1.02\pm0.04$ & $1.72\pm0.06$ \\
CO(3-2) & $5.5\pm1.6$ & $60.4\pm5.3$ & $10.4\pm1.7$ & $53.1\pm8.8$ & $0.83\pm0.15$ & $0.66\pm0.11$ \\
               & $-20.5\pm3.8$ & $155.8\pm7.7$ & $24.3\pm1.8$ & $124.1\pm9.1$ & $0.75\pm0.07$ & $1.5\pm0.1$ \\
$^{13}$CO(3-2) & $7.9\pm1.8$ & $84.5\pm4.3$ & $4.09\pm0.18$ & $22.8\pm1.0$ & $0.25\pm0.02$ & $0.285\pm0.013$ \\
\multicolumn{7}{l}{{\bf Primary clump}} \\
CO(3-2) & $24.8\pm1.6$ & $59.0\pm5.0$ & $4.9\pm0.7$ & $65.0\pm9.6$ & $1.04\pm0.18$ & $0.31\pm0.04$ \\
               & $-13.6\pm4.1$ & $152.0\pm5.7$ & $12.5\pm0.8$ & $165\pm11$ & $1.02\pm0.08$ & $0.80\pm0.05$ \\
$^{13}$CO(3-2) & $24.3\pm2.1$ & $86.9\pm5.0$ & $1.96\pm0.10$ & $28.4\pm1.4$ & $0.31\pm0.02$ & $0.136\pm0.007$ \\
\multicolumn{7}{l}{{\bf Secondary clump}} \\
CO(3-2) & $-4.7\pm1.1$ & $52.8\pm3.5$ & $6.4\pm0.7$ & $73.1\pm8.2$ & $1.3\pm0.2$ & $0.41\pm0.04$ \\
               & $-42.3\pm6.7$ & $142.5\pm9.3$ & $7.7\pm0.8$ & $87.9\pm9.3$ & $0.58\pm0.07$ & $0.49\pm0.05$ \\
$^{13}$CO(3-2) & $-1.8\pm1.4$ & $62.6\pm3.3$ & $1.78\pm0.08$ & $22.1\pm1.0$ & $0.33\pm0.02$ & $0.124\pm0.006$ \\
\hline
\hline
\end{tabular}
\end{center}
Notes: All linewidths have been corrected for instrumental broadening.
\label{tab:Ico}
\end{minipage}
\end{table*}

The full extent of the molecular gas is well-resolved for all observed spectral
lines. A spatially integrated spectrum was extracted for each spectral line 
from a $6.8''\times 5''$ ($13.6\times 10\kpc$) box encompassing all of 
the significant line emission. The CO(3-2) and $^{13}$CO(3-2) images were 
smoothed to match the resolution of the CO(1-0) image prior to the spectral 
extraction. Since the extraction region is much larger than either beam size, 
the smoothing has little impact on the spectrum. The spectra, shown in Fig. 
\ref{fig:isotopes}, are expressed as a brightness temperature:
\begin{equation}
  T_R = \frac{\lambda^2}{2k\Omega} S.
\end{equation}
Here $\lambda$ is the wavelength of the spectral line, $k$ is the Boltzmann 
constant, and $S$ is the measured flux density. For spatially resolved spectra, 
$\Omega$ is the solid angle of the spectral extraction region, provided it is 
much larger than the beam.

The spectra are all best fit by a single Gaussian velocity component. The 
$^{13}$CO(3-2) spectrum contains a second peak at $-80\kmps$, but the significance 
of the second component is below $3\sigma$, so has not been included in these 
fits. The results of the spectral fitting, including the integrated fluxes 
($S_{\rm CO}\Delta v$), integrated intensities ($W \equiv \int T_R {\rm d}v$), 
and peak temperatures ($T_{\rm peak}$), are listed in Table \ref{tab:Ico}. 
The peak temperature is related to the integrated intensity and the line 
full-width at half-maximum (FWHM) through $T_p=0.94 W_{\rm CO}/{\rm FWHM}$. 
The line luminosity, $L'_{\rm CO}$, is also included in Table \ref{tab:Ico}. 
It is defined as \citep{Solomon05} 
\begin{equation}
L'_{\rm CO} = 3.25\e{7} S_{\rm CO}\Delta v D_L^2 (1+z)^{-3} \nu_{\rm obs}^{-2} \Kkmpspcsq,
\end{equation}
where $S_{\rm CO}\Delta v$ is the flux density in $\Jy\kmps$, $D_L$ is the 
luminosity distance in $\Mpc$, and $\nu_{\rm obs}$ is the observed frequency
in GHz.

The total integrated CO(1-0) flux, $8.06\pm0.08\Jykmps$, is consistent within 
$2\sigma$ of the IRAM-30m single dish measurement of $9.9\pm1.0\Jykmps$ 
\citep{Edge01, Edge03}. Adopting the Galactic CO-to-H$_2$ conversion factor 
yields a molecular gas mass of $2\e{10}\Msun$.

\begin{figure}
  \centering
  \includegraphics[width=\columnwidth]{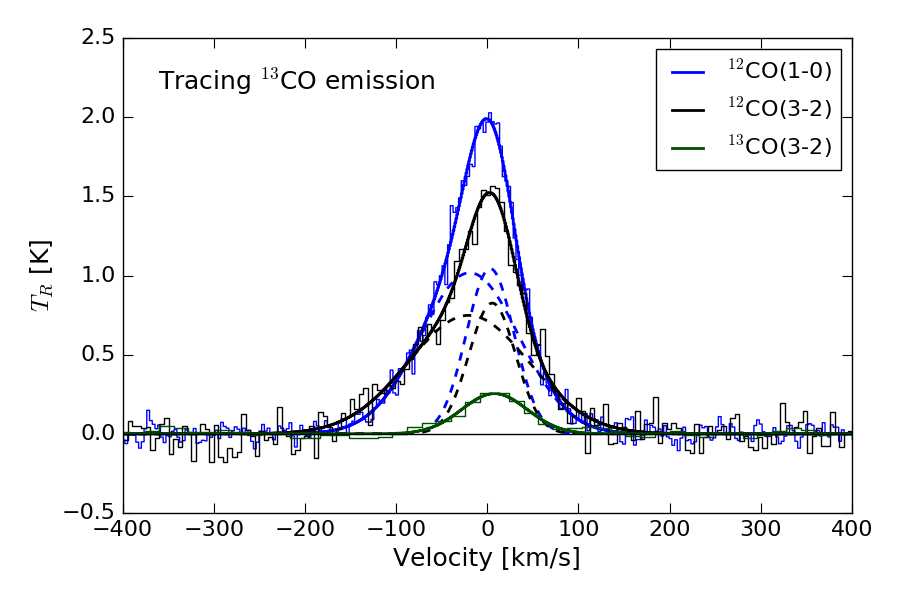}
  \caption{
    CO(1-0), CO(3-2) and $^{13}$CO(3-2) spectra extracted 
    from a region encompassing the two $^{13}$CO peaks seen in Fig. 
    \ref{fig:linemaps}. 
    The $^{12}$CO spectra were modelled with two Gaussian velocity 
    components, while the $^{13}$CO spectrum was modelled with a single 
    velocity component. The individual components for the $^{12}$CO 
    spectra are indicated by dashed lines.
  }
  \label{fig:polygonspectra}
\end{figure}

\begin{figure}
  \centering
  \includegraphics[width=\columnwidth]{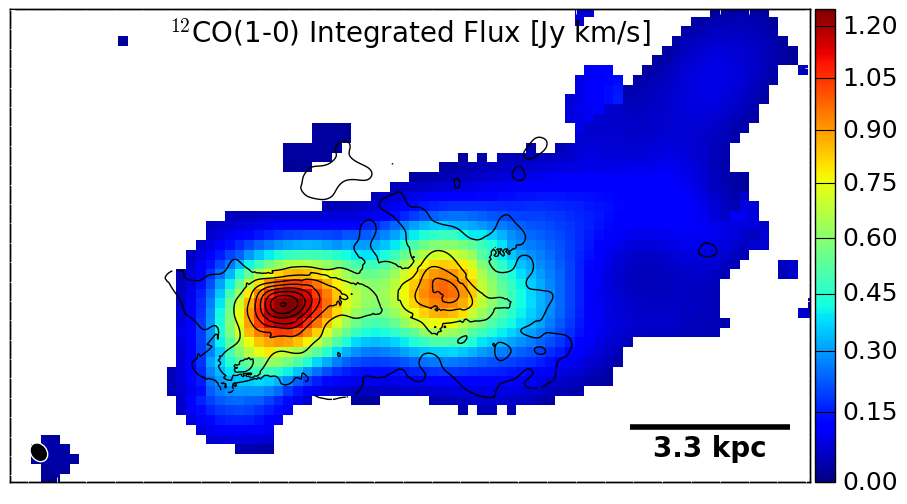}
  \includegraphics[width=\columnwidth]{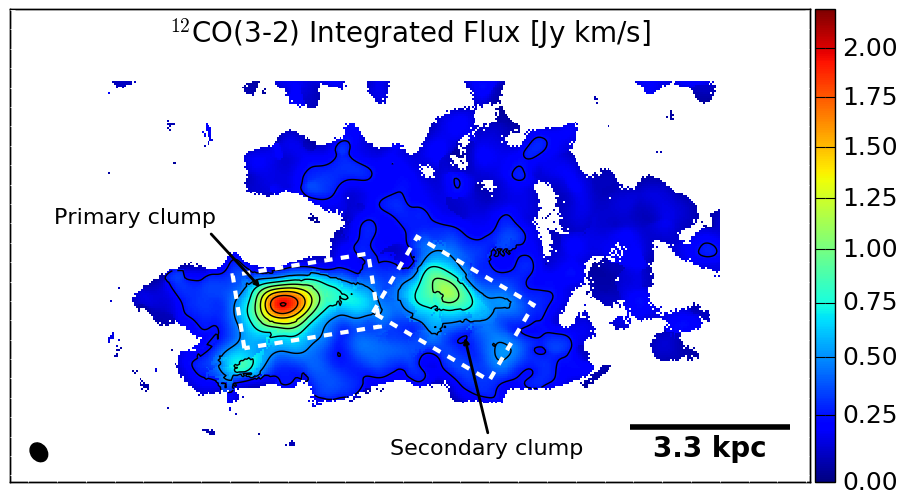}
  \includegraphics[width=\columnwidth]{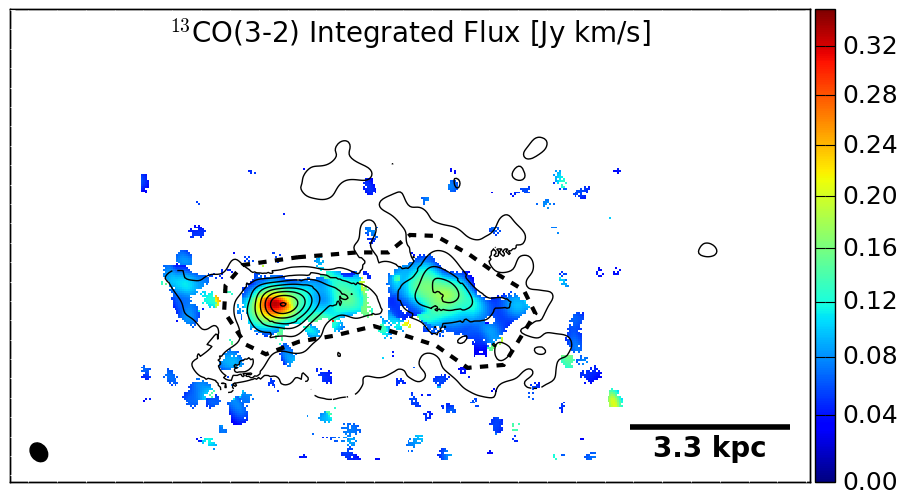}
  \caption{
    Integrated fluxes of the CO(1-0), CO(3-2), and $^{13}$CO(3-2) 
    transitions, in units of $\Jykmps$.
    Contours of CO(3-2) emission, ranging from $0.2$ to $1.96\Jykmps$ in 
    steps of $0.25\Jykmps$, have been overlaid on each map.
    The dashed lines in the center panel indicate the primary (left) and 
    secondary (right) clumps. The dashed line in the bottom panel indicates 
    the region tracing the $^{13}$CO emission used to extract the spectra for 
    Table \ref{tab:Ico}.
    The synthesized beam for each image is shown as the black ellipse in
    the lower left. 
    Each image shares the same $16.5\times 9.8\kpc$ field-of-view.
  }
  \label{fig:linemaps}
\end{figure}

In Table \ref{tab:Ico} we also include the fitted parameters for spectra 
extracted from a region tracing the $^{13}$CO(3-2) emission, which is distributed 
over smaller spatial scales than the $^{12}$CO emission (see Fig. 
\ref{fig:linemaps}). This ensures that molecular clouds detected only in 
$^{12}$CO are not biasing the comparisons between $^{12}$CO and $^{13}$CO.
The spectra are shown in Fig. \ref{fig:polygonspectra}. 
Note that the lower resolution of the CO(1-0) image results in some emission 
lying outside of the extraction region. The resolution of the CO(3-2) images 
have not been degraded to match the CO(1-0) resolution for these spectra. 

The velocity structure of the CO(1-0) and CO(3-2) lines are very 
similar. Both spectra have a narrow ($60\kmps$ FWHM) component at approximately 
$0\kmps$ in the adopted frame, as well as a broader ($130-160\kmps$) component 
that is blueshifted by about $20\kmps$. The linewidths of the narrow components 
are consistent, while the linewidth of the broader component is larger at 
CO(3-2) ($155.8\pm7.7\kmps$) than CO(1-0) ($127.2\pm2.2\kmps$).
The single peak seen in the $^{13}$CO spectrum, which is centered at $8\pm2\kmps$ 
with a linewidth of $84.5\pm4.3\kmps$, matches well with the narrow $^{12}$CO 
velocity component. No $^{13}$CO(3-2) emission corresponding to the broader 
$^{12}$CO line emission has been detected. We refer to these two velocity 
components seen in the $^{12}$CO spectra -- one with associated $^{13}$CO emission 
and one without -- as $^{13}$CO-bright and $^{13}$CO-faint.

\subsection{Gas Distribution}
\label{sec:distribution}

\begin{figure}
  \centering
  \includegraphics[width=\columnwidth]{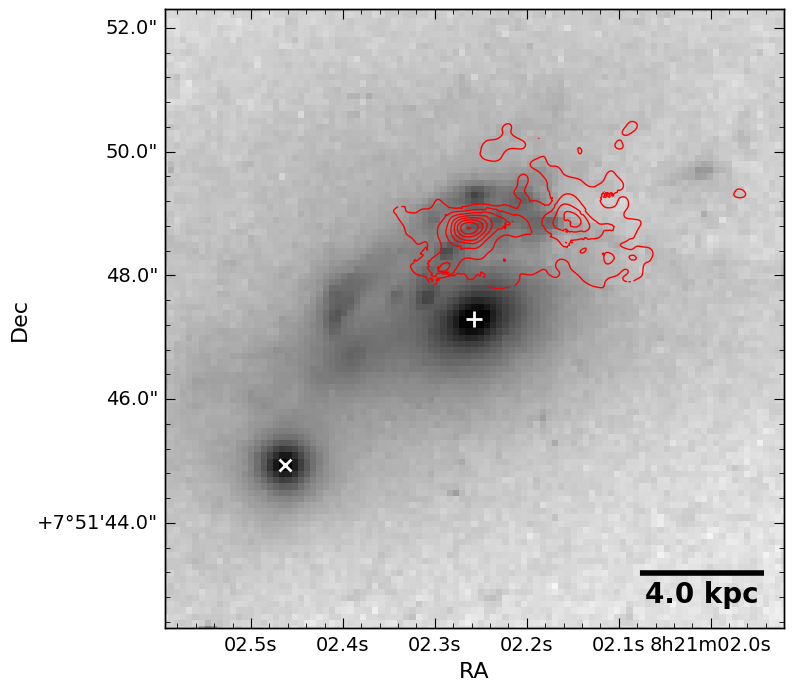}
  \caption{
    HST WFPC2 F606W imaging of the RXJ0821+0752 BCG overlaid with 
    contours of the CO(3-2) emission obtained from ALMA. The 
    $+$ indicates the BCG nucleus, and the $\times$ indicates the 
    centroid of a nearby galaxy that may be interacting with the BCG.
  }
  \label{fig:hst}
\end{figure}

Maps of integrated flux were created by fitting a one or two Gaussian components 
to the spectrum of each pixel after averaging over the size of the beam. 
The significance of the line was tested using a Monte Carlo simulation
with 1000 iterations, with a detection requiring a significance of $3\sigma$.
The integrated flux maps for the CO(1-0), CO(3-2), and $^{13}$CO(3-2) 
transitions are shown in Fig. \ref{fig:linemaps}. Contours of the CO(3-2) 
emission are overlaid on an {\it HST} WFPC2 F606W image in Fig. \ref{fig:hst}.

The molecular gas is distributed along an $\sim8\kpc$ long filament, with 
additional diffuse emission detected in the CO(1-0) map extending toward 
the northwest. Two clumps of molecular gas are seen in each map, and account 
for most of the emission. The primary clump, which accounts for $60\%$ of the 
total CO(1-0) flux, is located at the eastern end of the filament and is 
$3\kpc$ north of the BCG nucleus. The secondary clump is $3\kpc$ west of the 
primary clump. $^{13}$CO emission is only significantly detected near 
the brightest regions of the primary and secondary clumps. The molecular gas 
is closely associated with H$\alpha$ emission \citep{BayerKim02, Hamer16}.

We detect no significant emission centered on the nucleus of the BCG.  
This spatial offset will be discussed in a forthcoming paper alongside a 
recent {\it Chandra} observation.
The $3\sigma$ upper limits on integrated flux were computed following 
\citet{mcn94}, but using flux instead of intensity:
\begin{equation}
  S_{\rm CO}\Delta v = \frac{3\sigma_{\rm ch}\Delta V}{\sqrt{\Delta V/\Delta V_{\rm ch}}} \Jykmps.
\end{equation}
Here $\sigma_{\rm ch}$ is the RMS deviation in unsmoothed velocity channels 
in units of $\Jy$, $\Delta V$ is the expected full-width at half-maximum of 
the line, which we assume to be $300\kmps$, and $\Delta V_{\rm ch}$ is the 
velocity width of each channel. Extracting the CO(1-0) spectrum from a 
$2\kpc\times 2\kpc$ box centered on the optical centroid, we obtain a 
$3\sigma$ upper limit of $0.05\Jykmps$. This translates to a nuclear molecular 
gas mass $<1.2\e{8}\Msun$ assuming the Galactic CO-to-H$_2$ conversion factor. 

The optical imaging shows an arc of excess emission surrounding the northeastern 
portion of the BCG. The F606W filter covers the wavelength range $4309-6457\AA$ 
in the rest frame of the BCG, so does not include H$\alpha$ or [N{\sc ii}] emission.
The excess emission may originate from bluer emission lines, such as H$\beta$, 
or from stellar continuum tracing localized star formation or a tidal event 
involving a pre-existing stellar population. The arc is oriented toward a galaxy 
$7.7\kpc$ to the southeast of the BCG (SDSS J082102.46+075144.9), suggesting a 
possible interaction.

\citet{BayerKim02} obtained optical spectra of the BCG along two slits using 
the Intermediate dispersion Spectroscopic and Imaging System (ISIS) on the 
William Herschel Telescope (WHT). These slits encompassed two bright blobs along 
the arc located $1.6$ and $4.3\kpc$ east-northeast of the BCG nucleus. 
The blobs show elevated blue continuua compared to the surrounding stellar light. 
The innermost blob, which lies along the inner arc, is best-fitted by an older 
starburst with mainly A and F stars, while the blob along the outer arc shows a 
marginal O star component with significant amounts of all other stellar types. 
These spectra therefore show that the arc of excess emission is associated with 
recent star formation.

The primary clump of molecular gas and its diffuse extension to the southeast 
are coincident with small blobs located along the innermost arc. However, 
the outer arc, which contains the brightest of the optical blobs and extends 
toward the nearby galaxy, has little to no associated molecular gas. 
Additionally, the secondary clump extends toward a region with no enhanced 
optical emission.

\subsection{Line Ratios}
\label{sec:lineratios}

Important information about the physical state of the molecular gas is 
encoded in the ratio of brightness temperatures for each of the spectral lines.
The CO (3-2)/(1-0) brightness ratio is a function of the gas excitation 
temperature, $T_{\rm ex}$, and optical depth (see Eqn \ref{eqn:RJ}). 
The $^{12}$CO/$^{13}$CO (3-2) brightness ratio additionally depends on 
the relative isotopologue abundances.
Throughout this work we adopt the following notation to refer to 
these line ratios:
\begin{equation}
\begin{split}
r_{31} & \equiv T_{32}(^{12}{\rm CO})/T_{10}(^{12}{\rm CO}) \\
R_{10} & \equiv T_{10}(^{12}{\rm CO})/T_{10}(^{13}{\rm CO}) \\
R_{32} & \equiv T_{32}(^{12}{\rm CO})/T_{32}(^{13}{\rm CO}).
\end{split}
\label{eqn:r}
\end{equation}
In Table \ref{tab:lineratios} we list a number of published measurements 
of these ratios for comparison. Note that in some cases the published line 
ratios may be determined from integrated intensities (in units of $\Kkmps$) 
instead of brightness temperature. For two lines with the same linewidth, 
these are equivalent. We have opted to use brightness temperature over 
integrated intensity because brightness temperature is the quantity related 
directly to $T_{\rm ex}$ and optical depth, and the linewidth of the 
$^{13}$CO-faint component differs between CO(1-0) and CO(3-2).

\subsubsection{CO (3-2)/(1-0)}
\label{sec:r31}

From the peak temperatures listed in Table \ref{tab:Ico}, the ratio of 
CO (3-2)/(1-0) brightness temperatures from the spatially integrated 
spectra is $0.80\pm0.03$. Using the total integrated intensity instead gives a 
line ratio of $0.895\pm0.022$. The CO (3-2)/(1-0) intensity ratios in other 
BCGs that have been observed by ALMA are $0.8-0.9$ \citep{mcn14, Russell14, 
Russell16, Vantyghem16}. These are consistent with the ratio observed in 
RXJ0821, indicating that the excitation of the molecular gas is similar.

For the remaining discussion we restrict the line ratio measurements 
to the region tracing the $^{13}$CO emission. This is to more accurately 
reflect the excitation conditions of the $^{13}$CO-emitting gas, which is 
used to measure the molecular gas mass in Section \ref{sec:lte}. 
The CO(1-0) and CO(3-2) spectral fits for this region, shown in Table 
\ref{tab:Ico}, were measured at different resolutions. In order to compare 
these emission lines the CO(3-2) image was first smoothed to the resolution 
of the CO(1-0) image and the spectrum was re-extracted. The resulting 
$r_{31}$ line ratio is $0.71\pm0.15$ for the $^{13}$CO-bright velocity 
component and $0.65\pm0.06$ for the $^{13}$CO-faint component. The consistency 
between these two line ratios implies that the two components share similar 
gas excitation conditions. 

Individual molecular clouds in the Milky Way have mean densities of $300\pcmcu$ 
and temperatures of $10\K$, corresponding to a line ratio of $r_{31}=0.1-0.3$
\citep{Scoville87a}. 
The higher line ratio in RXJ0821 is indicative of enhanced molecular excitation, 
originating from molecular clouds with higher densities and/or temperatures. 
In external galaxies the intensity ratio is also higher than in individual 
molecular clouds. \citet{Mao10} measured $r_{31}$ in a sample of 125 nearby galaxies, 
subdividing the sample based on galaxy type. Their mean intensity ratios range 
from $0.61\pm0.16$ in normal galaxies to $0.96\pm0.14$ in (U)LIRGs. The $r_{31}$ 
in RXJ0821, which is derived from peak temperatures instead of integrated intensities, 
is consistent with the Seyfert and AGN-host galaxy populations. It is also consistent 
with the $r_{31}$ measured in H{\sc ii} regions \citep[e.g.][]{Wilson99}. The 
object-to-object scatter in $r_{31}$ is large, so it cannot be used to unambiguously 
infer excitation conditions based on galaxies with similar line ratios.

Assuming that the gas is thermalized and the emission is optically thick, 
the CO (3-2)/(1-0) line ratio can be used to estimate the gas temperature.
The Rayleigh-Jeans brightness temperature, $T_R$, for the $J\rightarrow J-1$ 
transition is given by:
\begin{equation}
  T_R = T_J \Phi_A [f(T_{\rm ex}) - f(T_{\rm bg})] (1 - {\rm e}^{-\tau_J}),
  \label{eqn:RJ}
\end{equation}
where $T_{\rm ex}$ is the excitation temperature of the transition, $\Phi_A$ 
is the area filling factor, $T_{\rm bg}=2.73(1+z)\K$ is the background 
temperature, $T_J = h\nu_J/k = T_1 J$ with $T_1=5.3\K$ for $^{13}$CO 
and $5.5\K$ for $^{12}$CO, and $f(T) \equiv (\exp(T_J/T)-1)^{-1}$. 
If the gas is in local thermodynamic equilibrium (LTE), then each transition 
will share a common excitation temperature that is equal to the gas kinetic 
temperature. Provided that the $^{12}$CO emission is optically thick, 
$1 - {\rm exp{(-\tau_J)}}\approx 1$ and the brightness temperature is 
independent of optical depth. We also assume that the CO(1-0) and CO(3-2) 
emission originate from the same regions in the molecular clouds, so the 
area filling factors are the same.

The ratio of CO (3-2)/(1-0) brightness temperatures for the $^{13}$CO-bright 
component, $0.71\pm0.14$, implies an excitation temperature of 
$T_{\rm ex}\approx 15\K$. Substituting this $T_{\rm ex}$ into Eqn. \ref{eqn:RJ} 
for the CO(3-2) line with its original resolution gives an area filling factor 
of $\Phi_A\approx 0.1$. The excitation temperature is a steep function of the 
line ratio, particularly when $r_{31}$ approaches unity. 
The $\pm1\sigma$ limits on $r_{31}$, $0.85$ and $0.57$, give excitation 
temperatures of $29\K$ and $9.3\K$, respectively.

An excitation temperature of $15\K$ is comparable to the low values observed 
in Galactic clouds. 
However, it is important to note that $T_{\rm ex}$ determined under the 
assumption of LTE is a lower limit to the actual kinetic temperature of the gas. 
The two temperatures are equal only when the gas densities exceed the critical 
density for each transition, which are $\sim10^{3}\pcmcu$ for CO(1-0) and 
$3\e{4}\pcmcu$ for CO(3-2). 
At lower densities the collision rate is too low to thermalize the gas. 
Reproducing the same line ratio with subthermal gas requires higher 
temperatures. 

\begin{table}
  \begin{minipage}{\columnwidth}
    \caption{Line Ratios in Other Galaxies}
	\begin{center}
	  \begin{tabular}{l c c c c}
	    \hline \hline
		Object   & $R_{10}$ & $R_{32}$ & $r_{31}$ & References \\
		\hline
		\multicolumn{4}{l}{{\bf RXJ0821.0+0752}} & This work \\
        All emission & -- & $7.6\pm1.5$ & $0.80\pm0.03$ & \\
        All emission$^{\dagger}$ & -- & $8.5\pm1.1$ & $0.895\pm0.022$ & \\
        Tracing $^{13}$CO$^{a}$ & -- & $3.32\pm0.66$ & $^{\ast}0.71\pm0.15$ & \\
        Tracing $^{13}$CO$^{b}$ & -- & $>14$ & $^{\ast}0.65\pm0.06$ & \\
		\hline
		\multicolumn{5}{l}{{\bf H{\sc ii} regions}} \\
		M17 & -- & $3.7\pm0.9$ & $0.76\pm0.19$ & 1 \\
		\hline
		\multicolumn{5}{l}{{\bf Starburst galaxies}} \\
		M82 & -- & $12.6\pm1.5$ & $0.8\pm0.2$ & 2, 3 \\
		NGC~253 & $11.5\pm1.9$ & $11.1\pm2.2$ & $1.08\pm0.18$ & 4 \\
		NGC~278 & $8.4\pm1.3$  & $11.4\pm1.7$ & 0.88 & 5 \\
		NGC~660 & $15.7\pm2.0$ & $12.8\pm1.9$ & 0.58 & 5 \\
		NGC~3628 & $12.2\pm1.8$ & $7.9\pm1.8$ & 0.44 & 5 \\
		NGC~4666 & $8.5\pm1.3$ & $11.3\pm1.7$ & 0.49 & 5 \\
		NGC~6946 & -- & $\sim7$ & $1.3$ & 6 \\
		\hline
		\multicolumn{5}{l}{{\bf (U)LIRGs and SMGs}} \\
		Arp 220  & $43\pm10$ & $8\pm2$ & $1.0\pm0.1$ & 7 \\
		NGC~6240 & $45\pm15$ & $>32$ & $\approx 1$ & 7 \\
		SMM~J2135 & $>31$ & $20\pm2$ & $0.68\pm0.03$ & 8 \\
		\hline \hline
	\end{tabular}
  \end{center}
	Notes: $^{\dagger}$Determined using integrated intensity instead of 
    peak temperature.
    $^{\ast}$Measured from the CO(3-2) image that has been smoothed 
    to match the resolution of the CO(1-0) image.
    $^{a}$The $^{13}$CO-bright velocity component.
    $^{b}$The $^{13}$CO-faint velocity component.
    References: 1: \citet{Wilson99}, 2: \citet{Petitpas00}, 3: \citet{Weiss05},
    4: \citet{Harrison99}, 5: \citet{Israel09}, 6: \citet{Wall93}, 7: \citet{Greve09},
    8: \citet{Danielson13}.
	\label{tab:lineratios}
	\end{minipage}
\end{table}

\subsubsection{$^{12}$CO/$^{13}$CO (3-2)}
\label{sec:R32}

Studies of the $^{12}$CO/$^{13}$CO intensity ratio are primarily conducted 
using the $J=1-0$ transition \citep[e.g.][]{Solomon79, Aalto95}. Individual 
giant molecular clouds have intensity ratios of $R_{10}=3-5$ \citep{Solomon79}. 
In external galaxies, the disks of spiral galaxies exhibit the lowest intensity 
ratios, with $R_{10}\approx6-8$. Intermediate ratios ($R_{10}\approx10-15$) are 
observed in starburst galaxies, while extreme ratios ($R_{10}>20$) are seen in 
merging systems \citep{Aalto95}.

If the $^{12}$CO and $^{13}$CO emission lines are both thermalized and originate 
from the same physical region in the molecular gas, then $R_{32}$ should be 
comparable to $R_{10}$. 
\citet{Israel09} measured $R_{10}$, $R_{21}$, and $R_{32}$ in a sample of five 
starburst galaxies, and found that $R_{32}$ and $R_{10}$ are consistent to within 
$50\%$ (see Table \ref{tab:lineratios}). The measured $R_{32}$ in starbursts range 
from roughly $8-13$.
Several merging systems also exhibit $R_{32}$ ratios that are comparable to 
$R_{10}$ \citep[e.g.][]{Greve09, Danielson13}. A notable exception is Arp~220, 
where $R_{10} = 43\pm10$ but $R_{32}=8\pm2$ \citep{Greve09}. This difference is 
attributed to a multi-component molecular gas distribution. The $^{13}$CO is 
primarily located in dense clumps while the bulk of the $^{12}$CO emission 
originates from an envelope of lower density gas with moderate optical depth 
($\tau\approx1$) \citep{Aalto95}.

Considering only the total integrated intensities derived from the 
spatially-integrated spectra, the $^{12}$CO/$^{13}$CO (3-2) line ratio 
is $8.5\pm1.1$. This global ratio is comparable 
to the lowest values observed in starburst galaxies. It is also consistent with 
the $R_{32}$ seen in Arp~220. However, the molecular gas in gas-rich mergers, 
such as Arp~220, is generally channeled into warmm, dense regions at the center 
of the galaxy, which is not the case in RXJ0821. Thus the global 
$^{12}$CO/$^{13}$CO (3-2) intensity ratio in RXJ0821 is most closely matched to the 
conditions of starburst galaxies. 

As noted in Section \ref{sec:spectra}, the $^{13}$CO emission is only associated 
with one of the two observed velocity components. Furthermore, the $^{13}$CO emission 
is confined to a much smaller spatial region than the $^{12}$CO emission.
In the $^{13}$CO-bright velocity component extracted from the region tracing the 
$^{13}$CO emission, the $^{12}$CO/$^{13}$CO (3-2) brightness temperature ratio is 
$3.32\pm0.66$. This $R_{32}$ is similar to those seen in H{\sc ii} regions 
\citep[e.g.][]{Wilson99}, as well as the $R_{10}$ in individual GMCs.

The other velocity component detected in $^{12}$CO shows no significant $^{13}$CO 
emission. With an RMS noise of $0.018\K$ in the $^{13}$CO(3-2) spectrum from Fig. 
\ref{fig:polygonspectra}, the 3$\sigma$ upper limit on peak $^{13}$CO(3-2) brightness 
temperature corresponds to a $^{12}$CO/$^{13}$CO (3-2) line ratio of $>14$.
The high $R_{32}$ in the $^{13}$CO-faint component implies either a higher 
$^{12}$CO/$^{13}$CO abundance ratio, a reduced optical depth, or extreme conditions 
from a gas-rich merger.

The isotopologue ratio is controlled by several proceses:
(i) $^{12}$C is produced primarily in massive stars, while $^{13}$C is a secondary 
product from a later stage of stellar processing. Young stellar ages would therefore 
enrich the interstellar medium with more $^{12}$C than $^{13}$C, leading to an 
increased [$^{12}$CO]/[$^{13}$CO] ratio. A top-heavy initial mass function would 
similarly favour $^{12}$C production over $^{13}$C.
(ii) The lower optical depth of $^{13}$CO is less effective at self-shielding 
from incident UV radiation. This can lead to selective photodissociation, 
where $^{13}$CO is photodissociated throughout a larger fraction of the cloud's 
volume than $^{12}$CO \citep{Bally82}.
(iii) In cold environments, ionized carbon atoms are exchanged with the CO 
isotopologues through chemical fractionation \citep{Watson76}:
\begin{equation}
^{13}C^+ + ^{12}CO \leftrightharpoons ^{13}CO + ^{12}C^+ + \Delta E,
\end{equation}
where $\Delta E=35\K$. For clouds below about $30\K$ the forward reaction 
is favoured and $^{12}$CO is converted into $^{13}$CO. In hotter environments 
the reaction reaches an equilibrium.
In starburst galaxies the elevated line ratios are attributed to the age 
of the stellar population. C$^{18}$O, which is produced alongside $^{12}$CO in 
young stars, is present at normal levels, and only the $^{13}$CO is depleted 
\citep[e.g.][]{Casoli92}. Early-type galaxies, on the other hand, show boosted 
$^{13}$CO/$^{12}$CO ratios due to their older stellar populations, which have 
had time to produce $^{13}$C in low-mass stars \citep{Alatalo15}.

A reduction in optical depth, assuming a constant isotopologue abundance 
ratio, leads to an increase in $R_{32}$. Since $\tau \propto N_{\rm CO}/\Delta v$, 
the $^{12}$CO optical depth is related to the $^{13}$CO optical depth simply 
through the $^{12}$CO/$^{13}$CO abundance ratio. Evaluating the optical depth 
from the ratio of $^{12}$CO to $^{13}$CO brightness temperatures (Eqn. \ref{eqn:RJ}), 
an $R_{32} > 14$ implies an $^{12}$CO optical depth $<3.5$, assuming 
$T_{\rm ex}=15\K$. Optically thin $^{12}$CO emission ($\tau < 1$) would require 
$R_{32}>31$.

In gas-rich mergers the $^{12}$CO/$^{13}$CO intensity ratio is elevated by 
turbulent motions introduced by the merger \citep[e.g.][]{Aalto95}. The 
turbulent motions increase the linewidth, decreasing optical depth and 
leading to an increase in the line ratio. Dissipation of this turbulence 
also heats the molecular gas. The combination of increased linewidth, higher 
temperatures, and selective photodissociation caused by the decreased  
$^{13}$CO optical depth leads to elevated $R_{32}$ in merging systems.
In RXJ0821 this is unlikely to be the case, as the narrow linewidths 
($<150\kmps$) are not indicative of turbulence introduced by a merger.

\subsubsection{Spatial Variation in Line Ratios}
\label{sec:spatialvariation}

To investigate the spatial variation of the molecular line ratios, we first 
extract spectra from the two main gas features: the primary and secondary clumps. 
Both CO(3-2) spectra exhibit the same velocity structure as the 
composite region tracing the $^{13}$CO emission -- a narrow peak is located at 
the systemic velocity and a broader component is blueshifted by about $30\kmps$ 
from the narrower peak. The CO(1-0) spectra were not extracted from these 
regions because the lower resolution moves a significant fraction of the flux 
outside of the region. 

The $R_{32}$ ratio for the $^{13}$CO-bright components of the primary and 
secondary clumps are $3.4\pm0.6$ and $3.9\pm0.7$, respectively. 
The main difference between these clumps is the significance of the broad, 
blueshifted wing. In the primary clump the wing accounts for $70\%$ of the 
total integrated flux, while in the secondary clump it accounts for only $55\%$. 
Computing $R_{32}$ in these clumps from the total intensity ratio, without 
isolating the $^{13}$CO-bright component, would lead to an artificially 
high value in the primary clump because of its large linewidth.
Instead, the peak temperature ratio of the $^{13}$CO-bright component is relatively 
constant between these two clumps.

\begin{figure}
  \centering
  \includegraphics[width=\columnwidth]{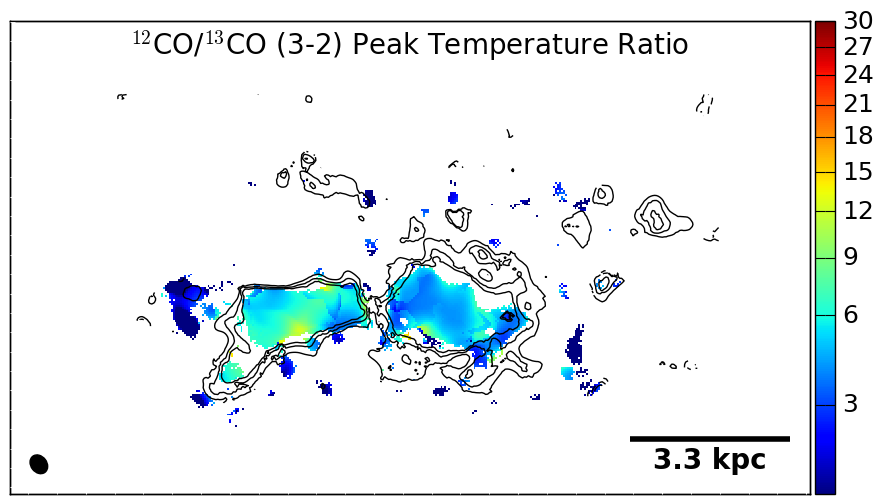}
  \includegraphics[width=\columnwidth]{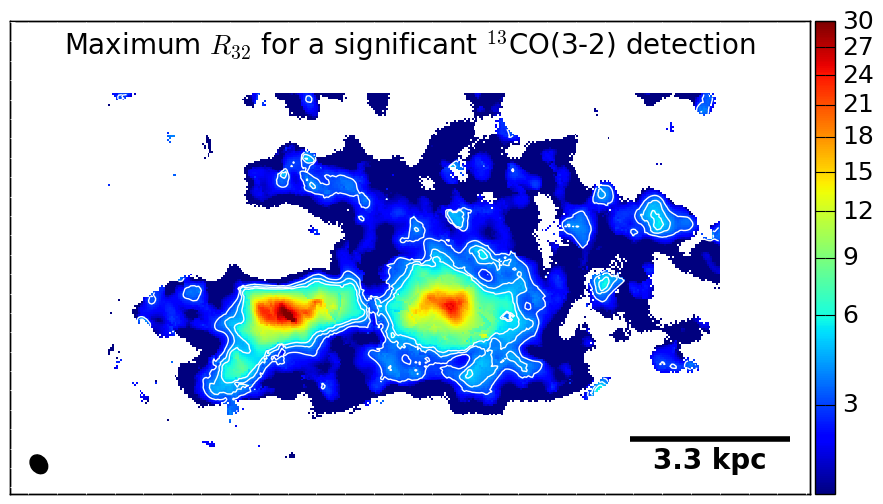}
  \caption{
    {\it Top:} Map of the $^{12}$CO/$^{13}$CO (3-2) brightness temperature 
    ratio. The $^{12}$CO emission has been modelled by two velocity components. 
    The component used in this map is the one with a velocity closest to 
    the $^{13}$CO emission.
    {\it Bottom:} The largest $R_{32}$ that would be detected in these observations, 
    determined from the ratio of $^{12}$CO brightness temperature to the $3\times$ 
    the RMS variation in the $^{13}T_{32}$ spectrum. Gas with an $R_{32}$ 
    below this threshold can be detected at $3\sigma$. 
    The contours in both images correspond to $R_{32,{\rm max}}$ of 3, 4, and 5.
  }
  \label{fig:R32}
\end{figure}

A map of the $^{12}$CO/$^{13}$CO (3-2) peak temperature ratio was created 
using the maps discussed in Section \ref{sec:distribution}. The temperature 
ratio was computed from the CO(3-2) velocity component that was closest 
in velocity to the $^{13}$CO(3-2) emission. 
When two velocity components are detected, this is generally the more 
redshifted peak. The map is shown in the top panel 
of Fig. \ref{fig:R32}. On the finer spatial scales used in the $R_{32}$ map, 
the line ratio in the primary clump is slightly elevated relative to the 
secondary clump. In the primary clump $R_{32}$ varies from roughly $4-7$, with 
an upward deviation to $12$ near the edge of the $^{13}$CO(3-2) detection. 
The secondary clump hosts an $R_{32}$ that ranges from $3.5-5$.

As the significantly detected $^{13}$CO(3-2) emission is confined to the 
regions that are brightest in CO(3-2), it is important to determine if 
$^{13}$CO throughout the rest of the gas 
is undetected because of an anomalously high $R_{32}$, or if an $R_{32}$ 
similar to the rest of the gas would also lie below our detection threshold.
To this effect we have created a map of the maximum 
$R_{32}$ that would be detectable given the observed $^{12}$CO brightness 
temperature and $^{13}$CO noise, $R_{32,{\rm max}} = ^{12}T_{32}/3^{13}T_{\rm rms}$.
This map is shown in Fig. \ref{fig:R32}. 
Since $^{12}$CO emission in the fainter regions cannot be uniquely attributed 
to the $^{13}$CO-bright component, we have computed $R_{32, {\rm max}}$ for 
both velocity components and taken the larger value. Gas with an $R_{32}$ below 
this threshold would have a peak $^{13}$CO temperature large enough to be 
detected by our imaging.

Overlaid on both images in Fig. \ref{fig:R32} are contours corresponding to an 
$R_{32,{\rm max}}$ of 3, 4, and 5. These are roughly in line with what has been 
detected throughout the rest of the gas. The primary and secondary clumps fill 
the majority of a contour of $R_{32,{\rm max}}=5$. Gaps within this contour may 
indicate slightly elevated line ratios. Outside of the two main clumps, very few 
regions reach an $R_{32,{\rm max}}$ of $5$. 
In these outer regions our observations are only sensitive to $^{13}$CO(3-2) 
intensities large enough to yield $R_{32}\lesssim 3$. As this is smaller 
than the line ratios observed in the rest of the gas, any $^{13}$CO located 
outside of the two main clumps lies below our detection threshold.

\section{CO-to-H$_2$ Conversion Factor}
\label{sec:lte}

The CO-to-H$_2$ conversion factor, $X_{\rm CO}$, relates the integrated 
intensity of a CO line to the H$_2$ column density (see Eqn. \ref{eqn:Xco}).
It can equivalently be expressed in terms of integrated properties:
\begin{equation}
  M_{\rm mol} = \alpha_{\rm CO} L'_{\rm CO},
\end{equation}
where the conversion factor $\alpha_{\rm CO}$ is now a mass-to-light 
ratio relating the total molecular gas mass (which includes a factor of 
1.36 to account for heavier elements) to the CO line luminosity in units 
of $\Kkmpspcsq$.
In the Milky Way and normal galaxies this conversion factor is 
$X_{\rm CO}=2\e{20}\pcmsq(\K\kmps)^{-1}$, or equivalently 
$\alpha_{\rm CO}=4.3 \Msun(\Kkmpspcsq)^{-1}$ \citep{Bolatto13}.

Here we use the $^{13}$CO(3-2) emission line to estimate $\alpha_{\rm CO}$ 
under the assumption of local thermodynamic equilibrium (LTE). 
We restrict this analysis to the spatial region encompassing the two bright 
clumps, which is shown in the bottom panel of Fig. \ref{fig:linemaps}. 
Additionally, only the $^{13}$CO-bright velocity component of the $^{12}$CO 
spectra from this region (Fig. \ref{fig:polygonspectra}) are considered. 
These restrictions ensure that the $^{12}$CO and $^{13}$CO considered originate 
from the same molecular gas. As a result, this LTE analysis is only sensitive 
to a fraction of the total molecular gas mass.

The total $^{13}$CO column density can be estimated from the intensity of 
a single transition $^{13}$CO($J\rightarrow J-1$) according to \citep{Mangum15}:
\begin{equation}
  N(^{13}{\rm CO}) = 
    \frac{3h}{8\pi^3\mu^2J} Q(T_{\rm ex}) 
    \frac{\exp (\frac{J+1}{2}h\nu_J/kT_{\rm ex}) }{ \exp (h\nu_J/kT_{\rm ex}) - 1}
    \tau_J \Delta v.
  \label{eqn:NCO}
\end{equation}
Here $\mu=0.11~{\rm Debye}=0.11\e{-18}~{\rm esu}$ is the $^{13}$CO dipole moment, 
$\frac{1}{2}(J+1)h\nu_J$ is the energy of rotational level $J$, $\nu_J$
is the frequency of the $J\rightarrow J-1$ transition, $\tau_J$ is the optical 
depth of the transition, and $\Delta v$ is the full linewidth.
Assuming that the population in each rotational level is described by a 
Boltzmann distribution at a common temperature $T_{\rm ex}$, the partition 
function is $Q(T_{\rm ex}) \approx 2T_{\rm ex}/T_1 + 1/3$, where 
$T_1\equiv h\nu_1/k=5.3\K$.
For the $^{13}$CO(3-2) transition, this expression becomes
\begin{equation}
  N(^{13}{\rm CO}) = 0.83\e{14} \pcmsq (T_{\rm ex} + 0.88)
    \frac{
        {\rm e}^{15.9\K/T_{\rm ex}} }{
        1 - {\rm e}^{-15.9\K/T_{\rm ex}}
    }
    \tau_J \Delta v.
  \label{eqn:NCOvals}
\end{equation}

In the limit of optically thin $^{13}$CO emission ($1-{\rm e}^{-\tau}\approx \tau$), 
the optical depth $\tau_J$ can be determined from Eqn. \ref{eqn:RJ}:
\begin{equation}
  \tau_J \approx 
    \frac{T_R}{  T_J \Phi_A [f(T_{\rm ex}) - f(T_{\rm bg})] }.
  \label{eqn:tau}
\end{equation}
This approximation begins to break down at $\tau \gtrsim 0.1$. In this regime 
the optical depth derived from Eqn. \ref{eqn:tau} must be multiplied by a factor 
of $\tau_J/(1-{\rm e}^{-\tau_J})$. From Section \ref{sec:r31}, the excitation 
temperature and area filling factor are $T_{\rm ex}=15\K$ and $\Phi_A=0.1$. 
The corresponding $^{13}$CO(3-2) optical depth obtained from Eqn. \ref{eqn:tau} 
is $\tau_3=0.30$, and the finite optical depth correction evaluates to 
$\approx 1.16$.

The optical depth can also be determined from the ratio of $^{12}$CO and 
$^{13}$CO brightness temperatures:
\begin{equation}
  \tau_J(^{13}{\rm CO}) \approx -\ln \left( 1 - 
    \frac{T_R(^{13}{\rm CO})}{T_R(^{12}{\rm CO})} \right).
  \label{eqn:tau_alt}
\end{equation}
Note that the dependence on $\Phi_A$ has reduced out of this expression, as 
it is assumed that both $^{12}$CO and $^{13}$CO originate from the same area.
This method yields an optical depth of $0.36$, which is consistent with 
Eqn \ref{eqn:tau} after the $\tau_J/(1-{\rm e}^{-\tau_J})$ correction.

The resulting column density in the optically thin limit, expressed now in 
terms of integrated intensity ($W_J = T_R/0.94~\Delta v$), is
\begin{multline}
  N_{\rm thin}(^{13}{\rm CO}) = 0.78\e{14} \pcmsq (T_{\rm ex} + 0.88)
    \frac{
        {\rm e}^{15.9\K/T_{\rm ex}} }{
        1 - {\rm e}^{-15.9\K/T_{\rm ex}}
    } \\ \times
    \frac{^{13}W_{32} }{
        T_J \Phi_A [f(T_{\rm ex}) - f(T_{\rm bg})]
    },
  \label{eqn:NCOfinal}
\end{multline}
and the corrected optical depth is $N = N_{\rm thin} \times
\tau_J/(1-{\rm e}^{-\tau_J})$.
For $T_{\rm ex}=15\K$, $\Phi_A=0.1$, and $^{13}W_{32}=22.8\Kkmps$, the total 
$^{13}$CO column density is $1.74\e{17}\pcmsq$.
Assuming the abundance ratios $[^{12}{\rm CO}]/[^{13}{\rm CO}] = 50$ and 
$[^{12}{\rm CO}]/[{\rm H}_2] = 10^{-4}$ \citep{Dickman78, Frerking82} yields an 
H$_2$ column density of $N_{\rm H_2}=8.7\e{22}\pcmsq$. 
The total molecular gas mass is then computed from 
$M_{\rm mol} = 1.36 m_{\rm H_2} N_{\rm H_2} A_{\rm source}$, where 
$A_{\rm source}=\Phi_A A_{\rm reg}$ is the surface area of the molecular gas and 
$A_{\rm reg}=11.2\kpc^2$ is the area of the region used to extract the spectra. 
This gives a total molecular gas mass of $2.1\e{9}\Msun$. 

The CO-to-H$_2$ conversion factor is calibrated using CO(1-0) luminosity. Since 
some CO(1-0) emission is spread outside of this region from its lower resolution, 
it is estimated from the CO(3-2) luminosity ($0.66\e{9}\Kkmpspcsq$; see Tab. 
\ref{tab:Ico}) assuming a constant $r_{31}$ of $0.71$. 
The inferred CO(1-0) luminosity for the $^{13}$CO-bright velocity component is 
$L'_{10} = L'_{32}/r_{31} = 9.3\e{8}\Kkmpspcsq$, giving a CO-to-H$_2$ conversion 
factor of $\alpha_{\rm CO} = 2.26\Msun(\Kkmpspcsq)^{-1}$. This is half of the 
Galactic value.

The total luminosity of both CO(1-0) components from within the region 
tracing the $^{13}$CO emission, assuming an $r_{31}$ of $0.7$, is $3\e{9}\Kkmpspcsq$. 
Adopting this total luminosity in place of that from the $^{13}$CO-bright 
component only yields a CO-to-H$_2$ conversion factor of $\alpha_{\rm CO} = 
0.7\acounits$, which is $6$ times lower than the Galactic value. This is 
comparable to the standard value of $\alpha_{\rm CO}=0.8\acounits$ in ULIRGs 
and starburst galaxies \citep{Downes98, Bolatto13}. 
However, this conversion factor uses the molecular gas mass that was measured 
from the single $^{13}$CO velocity component, whereas two distinct components 
were detected at $^{12}$CO. Any mass in the $^{13}$CO-faint velocity component 
has therefore been neglected in this measurement.

In order to investigate the dependence of the molecular gas mass measurement 
on the spatial region used to extract the spectra, we repeat this analysis for the 
region containing all of the line emission. The line ratio throughout the entire 
gas distribution is $r_{31}=0.80\pm0.03$, corresponding to $T_{\rm ex}=22\K$ 
and $\Phi_A=0.0126$. 
The $^{13}$CO(3-2) optical depth computed from Eqn. \ref{eqn:tau_alt} is $0.14$. 
The total $^{13}$CO column density is $1.3\e{17}\pcmsq$, which is slightly 
lower than the column density obtained from the smaller region. This implies 
a molecular gas mass of $2.47\e{9}\Msun$ over the $13.6\times 10\kpc$ box 
used to obtain the spectra. The corresponding CO-to-H$_2$ conversion factor, 
using the full CO(1-0) luminosity, is $0.53\Msun(\Kkmpspcsq)^{-1}$, in close 
agreement with the smaller region when both velocity components are included 
in the line luminosity. From these measurements, about $85\%$ of the molecular 
gas mass traced by $^{13}$CO is contained within the primary and secondary clumps.

The absence of $^{13}$CO emission from the broad, blueshifted wing implies 
that the measured molecular gas mass does not trace the entire supply of 
molecular gas. The molecular gas mass that is traced by the $^{13}$CO-bright 
velocity component is $2.1\e{9}\Msun$. 
Assuming that the CO-to-H$_2$ conversion factor in this parcel of gas is the 
same as the remainder of the gas, the total CO(1-0) luminosity gives a total 
molecular gas mass of $1.1\e{10}\Msun$.

\subsection{Underlying Assumptions}
\label{sec:uncertainties}

LTE models are highly simplistic in nature. Several assumptions, both direct 
and indirect, have been required to enable this analysis. Here we discuss the 
major assumptions, and comment on how they may affect our results.

\subsubsection{$^{13}$CO Abundance}

Without a direct probe of the H$_2$ content of the molecular gas, converting 
the measured $^{13}$CO column density into an H$_2$ column density requires 
the assumption of a $^{13}$CO abundance. We assumed a CO abundance of 
$[^{12}{\rm CO}]/[{\rm H}_2] = 10^{-4}$ and an isotopologue abundance 
ratio of $[^{12}{\rm CO}]/[^{13}{\rm CO}] = 50$. The estimated CO-to-H$_2$ 
conversion factor depends linearly on the assumed $[^{13}{\rm CO}]/[{\rm H}_2]$. 
In many extragalactic observations these quantities are not measured 
directly, so these assumptions are common.

For cloud metallicities above $\sim0.1\Zsun$ the CO/H$_2$ abundance ratio varies 
linearly with metallicity \citep{Bialy15}. 
In cool core clusters the molecular gas is likely formed from the cooling of the 
hot atmosphere, so the cloud abundance should be related to the metallicity of 
the hot atmosphere. In RXJ0821 the metallicity is $\sim0.4\Zsun$ within the 
central $10\kpc$ of the cluster core, and peaks at $\sim0.8\Zsun$ at about 
$30\kpc$ (\citealt{BayerKim02}; Vantyghem et al. in prep). 
If the abundance in the molecular clouds reflects the central metallicity, then 
the CO/H$_2$ abundance ratio would be overestimatedand, and the CO-to-H$_2$ 
conversion factor underestimated, by a factor of $2.5$. A declining metallicity 
profile is also observed in Perseus, where \citet{Panagoulia13} suggested that 
the missing metals near the core are locked up in cold dust. The molecular clouds 
may therefore be more metal-rich than the central atmosphere, implying an abundance 
ratio closer to solar.

The isotopologue ratio of 50 was chosen to represent the midpoint of this observed 
range in the Milky Way, which increases radially from 24 in the Galactic Center to 
$>100$ at large radii \citep{Langer90, Milam05}. Variations in the isotopologue ratio 
are controlled by stellar processing, selective photodissociation, and chemical 
fractionation (see Section \ref{sec:R32}). The data available for extragalactic 
sources seems to indicate that the isotopologue ratio increase with redshift, 
with values of $\sim40$ in local starbursts and $100$ or higher in ULIRGs \citep{
Henkel14}. An elevated isotopologue ratio in RXJ0821 would bring the measured 
conversion factor closer to the Galactic value. Reconciling the two would require 
$^{12}$CO/$^{13}$CO~$\approx 100$.

\subsubsection{$^{13}$CO Emitting Area}

Throughout this analysis we have assumed that the $^{12}$CO and $^{13}$CO 
emission originate from the same emitting area, thus sharing a common area 
filling factor. This assumption can break down in two ways. First, differences 
between the $^{12}$CO and $^{13}$CO optical depths can lead to selective 
photodissociation of $^{13}$CO in the lower density outskirts of individual 
molecular clouds. As a result, $^{13}$CO will be confined to the central regions 
of the clouds, while $^{12}$CO extends throughout the cloud and is well-mixed 
with H$_2$. This would decrease the $^{13}$CO filling factor, and the measured 
optical depth must be modified by the ratio of the areas. The true $^{13}$CO 
optical depth would then be underestimated in our analysis, leading to an 
underestimate in the conversion factor. For significant selective photodissociation, 
the $^{13}$CO emission will originate from a portion of the cloud with different 
physical conditions than the $^{12}$CO emission. The temperature, density, and 
column density for the gas containing $^{12}$CO and $^{13}$CO would need to be 
measured independently.

Second, the $^{12}$CO and $^{13}$CO may be distributed over different spatial 
regions in the entire cloud ensemble. In these observations $^{13}$CO is 
detected only in the two bright clumps, while diffuse $^{12}$CO encompasses 
a larger envelope. This may indicate that the diffuse emission contains little 
$^{13}$CO, from little stellar processing, significant selective photodissociation, 
or other processes that enhance the $^{12}$CO/$^{13}$CO abundance ratio. 
However, as shown in Section \ref{sec:spatialvariation}, we can only detect 
$^{13}$CO outside of the two main clumps if the $^{13}$CO abundance is abnormally 
large. By restricing the LTE analysis to the region tracing the $^{13}$CO 
emission (see Fig. \ref{fig:linemaps}, bottom), we ensure that the $^{12}$CO 
and $^{13}$CO emission trace the same cloud ensemble.

\subsubsection{Thermal Equilibrium}

The fundamental assumption of the LTE model is that each rotational transition 
can be described by a common excitation temperature that is equal to the kinetic 
temperature of the molecular gas. This approximation is only satisfied if the 
density of the molecular gas exceeds the critical density for the transition, 
which is $\sim10^{3}\pcmcu$ for CO(1-0) and $3\e{4}\pcmcu$ for CO(3-2). In this 
regime, collisions with H$_2$ occur rapidly enough to thermalize the CO molecules.

The CO (3-2)/(1-0) peak temperature ratio of $0.71$ for the $^{13}$CO-bright 
component implies thermalized emission only if the gas temperature is $15\K$. 
Alternatively, the gas may be hotter and subthermally excited. 
Subthermal excitation of the CO(3-2) line modeled under LTE conditions can lead 
to an underestimate in column density. For example, \citet{Nishimura15} performed 
an LTE analysis using both $^{13}$CO(1-0) and $^{13}$CO(2-1) in the Orion giant 
molecular cloud. The column density derived using $^{13}$CO(2-1) was a factor of 
three lower than that derived using $^{13}$CO(1-0), which they attribute to 
subthermal excitation.

Within the central $10\kpc$ of the RXJ0821 galaxy cluster, the pressure 
of the hot atmosphere is $2\e{-10}\ergpcmcu$ (Hogan et al. priv comm). This 
confining pressure sets the minimum pressure of the molecular clouds. If the 
gas is in pressure balance, then a kinetic temperature of $15\K$ corresponds 
to a density of $9\e{4}\pcmcu$, which exceeds the CO(3-2) critical density 
by a factor of three. The molecular gas should then be thermalized for any 
temperature up to $\sim50\K$. Additionally, the pressure in self-gravitating 
clouds exceeds the confining pressure, indicating that the molecular gas in 
this system is likely thermalized.

\section{Discussion}

In the subset of the molecular gas traced by $^{13}$CO emission, the 
CO-to-H$_2$ conversion factor is $\alpha_{\rm CO}=2.26\acounits$, or 
equivalently $X_{\rm CO}=1.04\e{20}\pcmsq(\Kkmps)^{-1}$. This is the 
first measurement of $X_{\rm CO}$ in a BCG. Previous works have simply 
adopted the Galactic value, justifying the decision based on the high 
metal abundances at the centers of galaxy clusters and the low expected 
temperature of gas condensing from the hot atmosphere. Our results 
indicate that $X_{\rm CO}$ in RXJ0821 is half of the Galactic value. 
Given the high scatter in extragalactic determinations of $X_{\rm CO}$, 
this measurement is broadly consistent with the Galactic value. Continuing 
to adopt the Galactic value in other BCGs may lead to an overestimate 
of $M_{\rm mol}$, but is likely accurate to within a factor of two.

Accurate measurements of the molecular gas mass in BCGs are crucial in 
understanding the gas origin and its role in AGN feedback. In particular, 
molecular flows trailing X-ray cavities have been detected in several 
BCGs \citep{Salome11, mcn14, Russell16, Russell17, Vantyghem16}, with the 
cold gas either lifted directly by the cavities or has cooled in situ from 
uplifted, hot gas. By Archimedes' principle, the maximum mass of uplifted 
gas is limited by the mass displaced by the X-ray cavities. Each of the 
observed outflows requires a high coupling efficiency between the cavities 
and uplifted gas, with the displaced mass exceeding the uplifted molecular 
gas mass by factors of a few. For example, $10^{10}\Msun$ of cold gas in A1835 
trails the X-ray cavities, which have displaced $3\e{10}\Msun$ of hot gas 
\citep{mcn14}. Reducing the CO-to-H$_2$ conversion factor alleviates the 
requirement of a high coupling efficiency.

In jet-driven molecular outflows, the outflowing gas may become optically 
thin while the gas in the disk remains optically thick \citep{Dasyra16}. 
In RXJ0821 the $^{12}$CO/$^{13}$CO line ratios in the $^{13}$CO-bright component 
imply $^{12}$CO optical depths $\gg1$, indicating that its molecular gas, which 
is spatially offset from the galactic nucleus, does not resemble the gas in 
jet-driven outflows. The $^{13}$CO-faint component is also consistent with 
moderate $^{12}$CO optical depths. Extending these results to the gas flows 
observed in other BCGs implies a gentler lifting process than in fast, 
jet-driven outflows.

\section{Conclusions}
\label{sec:conclusions}

We have presented new ALMA cycle 4 observations of the CO(1-0), CO(3-2), and 
$^{13}$CO(3-2) emission lines in the BCG of the cool core cluster RXJ0821+0752. 
The is the first detailed study of a $^{13}$CO emission line in a BCG. 
We have used the optically thin $^{13}$CO emission to estimate the molecular 
gas mass without relying on the Galactic CO-to-H$_2$ conversion factor. 
Our results are summarized as follows:
\begin{itemize}
  \item The molecular gas is primarily situated in two clumps located $3\kpc$ 
north to northwest of the galactic nucleus, with $<1.2\e{8}\Msun$ coincident 
with the BCG nucleus. These clumps and their surrounding diffuse emission are 
part of an $8\kpc$ long filament. $^{13}$CO(3-2) emission is only detected within 
the two bright clumps. Any $^{13}$CO emission located outside of these clumps 
lies below our detection threshold.
  \item Both of the $^{12}$CO spectra extracted from a region tracing the 
$^{13}$CO emission contain two velocity components. The narrower component
($\sim60\kmps$ FWHM) is consistent in both velocity centroid and linewidth 
with the $^{13}$CO(3-2) emission. The broader ($130-160\kmps$), slightly 
blueshifted ($\sim30\kmps$) wing has no associated $^{13}$CO(3-2) emission.
  \item Assuming that the molecular gas is in local thermodynamic equilibrium 
at a temperature of $15\K$, the molecular gas mass traced by the $^{13}$CO 
emission is $2.1\e{9}\Msun$. Isolating the $^{12}$CO velocity component that 
accompanies the $^{13}$CO emission yields a CO-to-H$_2$ conversion factor of 
$\alpha_{\rm CO}=2.26\acounits$, which is a factor of two lower than the 
Galactic value. 
\end{itemize}
Adopting the Galactic conversion factor in BCGs, as is currently the common 
practice, may lead to slight overestimates of $M_{\rm mol}$. However, 
the factor of two difference between the the measured and Galactic CO-to-H$_2$ 
conversion factors is comparable to object-to-object variations \citep{Bolatto13}. 
Continuing to adopt the Galactic conversion factor in BCGs should be reasonable 
until this analysis, or a complete excitation analysis, can be conducted in 
other systems.

\acknowledgements

We thank the anonymous referee for helpful comments that improved the paper.
Support for this work was provided in part by the National Aeronautics and Space Administration through Chandra Award Number G05-16134X issued by the Chandra X-ray Observatory Center, which is operated by the Smithsonian Astrophysical Observatory for and on behalf of the National Aeronautics Space Administration under contract NAS8-03060.
ANV and BRM acknowledge support from the Natural Sciences and Engineering Research Council of Canada.
BRM further acknowledges support from the Canadian Space Agency Space Science Enhancement Program.
ACE acknowledges support from STFC grant ST/P00541/1.
ACF and HRR acknowledge support from ERC Advanced Grant Feedback 340442.
This paper makes use of the following ALMA data: ADS/JAO.ALMA 2011.0.00735.S, 2012.1.00988.S, and 2016.1.01269.S. ALMA is a partnership of the ESO (representing its member states), NSF (USA) and NINS (Japan), together with NRC (Canada), NSC and ASIAA (Taiwan), and KASI (Republic of Korea), in cooperation with the Republic of Chile. The Joint ALMA Observatory is operated by ESO, AUI/NRAO, and NAOJ.
This research made use of APLpy, an open-source plotting package for Python hosted at http://aplpy.github.com.

\bibliographystyle{apj}
\bibliography{RXJ0821_13CO}

\end{document}